\documentclass[mlmain]{jmlr}

\usepackage{booktabs}
\usepackage{multirow}

\jmlrvolume{}
\firstpageno{1}

\jmlryear{2026}
\jmlrworkshop{Symmetry and Geometry in Neural Representations}

\title[Alignment De-Identifies Without Aligning]{Alignment De-Identifies Without Aligning: A Congruence-Invariant Relational Gap in Cross-Subject Neural Representations}

\author{
\Name{Anna Notaro} \Email{anna.notaro@student.uni-tuebingen.de}\\
\addr University of T\"{u}bingen, T\"{u}bingen, Germany\\
\addr Konrad Zuse School of Excellence in Learning and Intelligent Systems (ELIZA), Germany
\AND
\Name{Letizia Rossato} \Email{letro@dtu.dk}\\
\Name{Sadasivan Puthusserypady} \Email{sapu@dtu.dk}\\
\addr Department of Health Technology, Technical University of Denmark, Kongens Lyngby, Denmark
}
\begin{document}

\maketitle

\begin{abstract}
In brain-computer interfaces (BCIs), inter-subject variability remains a major challenge that limits decoding performance across users. Subject-wise alignment is widely used to improve cross-subject transfer of neural decoders, yet it is typically evaluated only through decoding accuracy, which cannot determine whether aligned representations share greater task-related structure. We therefore distinguish two questions that accuracy conflates: whether alignment removes subject-specific information and whether it makes the relational organization of task conditions more similar across individuals. Using three motor imagery (MI) electroencephalogram (EEG) cohorts, trials are represented as covariance matrices and aligned by recentering each subject's mean covariance to a common reference. Subject-specific information is quantified with a subject-identification probe, whereas relational structure is assessed using representational dissimilarity matrices (RDMs) computed from condition means with the affine-invariant Riemannian distance. We show theoretically that recentering leaves this distance, and therefore the resulting RDMs, unchanged. Consistent with this result, recentering reduces subject identification accuracy from 0.96 to chance level across all cohorts, while altering RDMs only at numerical precision. Cross-subject agreement in relational structure is significantly above zero but below the estimator's lower noise ceiling. Introducing an explicit similarity regularizer during training also fails to increase held-out agreement: the apparent reduction in between-subject differences is explained by a decline in the reliability ceiling rather than improved alignment. These findings show, in practice as well as in principle, that eliminating subject identity is not equivalent to aligning relational geometry. More broadly, alignment claims must be interpreted relative to the invariances of both the distance and the statistic used to evaluate them.
\end{abstract}

\begin{keywords}
Representational Similarity Analysis (RSA), Representation Geometry, Invariance, Riemannian Geometry, Brain-Computer Interface (BCI), Cross-Subject Alignment, Electroencephalogram (EEG).
\end{keywords}

\section{Introduction}
\label{sec:intro}

In brain-computer interface (BCI) systems, recorded neural activity is decoded into control signals. Among non-invasive recording modalities, electroencephalogram (EEG) is the most commonly employed because it is safe, inexpensive, portable, and readily reusable. Motor imagery (MI), in which a person imagines a movement without executing it, is among the most widely used paradigms: imagined movement of the hands, feet, or tongue modulates sensorimotor rhythms in a way a decoder can map to discrete commands \citep{barachant2012,congedo2017}. In post-stroke neurorehabilitation, a MI-BCI can detect an imagined movement and trigger contingent feedback, reinstating a sensorimotor loop that the lesion has interrupted, and improving motor outcomes in subacute stroke \citep{pichiorri2015}.

The clinical value of such a system depends on its ability to generalize to new patients. This requirement is hard to meet because MI EEG varies strongly between subjects, depending on skull and cortical anatomy, electrode placement, skin impedance, and the adopted imagery strategy. The resulting shift in EEG features distribution is the main obstacle to transferring a decoder between subjects, especially between stroke patients with lesions of different size and location \citep{saha2020}. Recovering accuracy with per-subject calibration costs time and fatigue to patients in rehabilitation. The standard remedy is per-subject alignment: mapping each subject to a common reference before decoding \citep{he2020,zanini2018,kobler2022}. Its success, however, is almost always assessed through decoding accuracy, which provides no insight about how alignment has altered the internal geometry of the representation.

This work reflects upon what per-subject alignment changes in that geometry, separating two properties that decoding accuracy conflates. \emph{Subject identity} is the extent to which a single trial reveals which subject produced it. \emph{Relational structure} refers to the arrangement of task conditions relative to one another within a subject, capturing which imagined movements are represented as similar and which as distinct. These properties are independent: recentering moves every subject's mean covariance onto a common reference, not modifying the configuration of their condition means around that center. Two subjects can therefore be brought to the same location while still retaining a different arrangement for the same imagined movements. Removing the information that identifies a subject is consequently not the same operation as making subjects' condition geometries agree. This reasoning explains why the two properties should therefore be evaluated separately when studying cross-subject decoding.

This study makes four major contributions:
\begin{enumerate}
\item It shows that per-subject alignment removes subject identity while leaving measured cross-subject relational agreement unchanged, so de-identification does not by itself deliver relational alignment (\sectionref{sec:dissociation}).

\item It explains this result through a known invariance. The representational dissimilarity matrix (RDM) is built from the affine-invariant Riemannian (AIRM) distance, under which per-subject congruences such as recentering are isometries \citep{zanini2018,he2020}. Agreement, ceiling, and their difference are preserved exactly and the measured change is at floating-point level
(\theoremref{thm:invariance}).

\item It shows that optimizing directly for cross-subject similarity during training, using a Centered Kernel Alignment (CKA)-based loss term, does not improve held-out agreement: decrease in difference is due to falling reliability ceiling rather than rising agreement (\sectionref{sec:training}).

\item It shows that sharedness is not why the tested maps leave the RDM unchanged, by constructing a single transformation applied to all subjects that drives agreement to its maximum on a finite sample, narrowing the open question to shared maps of bounded flexibility that must generalize to held-out trials (\sectionref{sec:sharedness}).
\end{enumerate}

The practical message is that the per-subject alignment acts differently on two components of
inter-subject variability. It removes distributional offsets that make the trial's subject
identifiable and, in the tested settings, also improves cross-subject decoding, while leaving
the AIRM condition geometry unchanged.  More generally, a relational measure cannot register an alignment that lies within its invariance class, which is set jointly by the distance building the RDM and the statistic comparing RDMs. Such an invariance result is therefore a property of the measurement, not evidence about alignment.

\section{Related work and theoretical background}
\label{sec:background}

\paragraph{Covariance representations on the SPD manifold.}
A standard representation for oscillatory EEG summarizes each trial by its spatial covariance across channels, which captures the band-power and inter-channel structure that MI modulates. Covariance matrices are symmetric and positive definite (SPD), so they lie on a curved manifold $\mathcal{P}_d$ of $d \times d$ SPD matrices. Treating them as flat vectors distorts the distances between the matrices. The affine-invariant Riemannian metric (AIRM) is the distance
\begin{equation}
\delta(\mathbf{P},\mathbf{Q})=\left\|\log\left(\mathbf{P}^{-1/2}\mathbf{Q}\mathbf{P}^{-1/2}\right)\right\|_F ,
\end{equation}
where $\mathbf{P}$ and $\mathbf{Q}$ are any two SPD matrices and $\|\cdot\|_F$ is the Frobenius norm. Classifiers built on this geometry, either by nearest Riemannian mean or by projection to the tangent space, are established baselines for MI decoding
\citep{barachant2012,congedo2017}; this work adopts that representation and studies its changes related to alignment.

\paragraph{Alignment by recentering.}
The dominant source of cross-subject mismatch in this representation is that each subject's covariances are centered at a different point of the manifold \citep{saha2020}. Two families of methods can correct for this variability. \emph{Global} recentering estimates one reference from the pooled group of subjects and applies it to everyone, changing the coordinate system but leaving the displacement between subject means intact. \emph{Per-subject} recentering estimates a separate reference for each subject and maps each subject's own mean onto a common point, removing differences between subjects. Euclidean alignment \citep{he2020} and Riemannian recentering \citep{zanini2018} are the standard instances of the second family, while TSMNet learns a related per-domain operation inside the network through SPD domain-specific batch normalization (SPDDSMBN) \citep{kobler2022}.

\paragraph{Comparing representations.}
Representational Similarity Analysis (RSA) compares systems by the matrix of pairwise dissimilarities between conditions, representational dissimilarity matrix (RDM), which makes comparison possible across systems with no common coordinates \citep{kriegeskorte2008,nili2014}. Related measures such as Centered Kernel Alignment (CKA) \citep{kornblith2019} and generalized shape metrics \citep{williams2021} pursue the same goal under different invariance assumptions. Because RDMs estimated from finite noisy data are themselves unreliable, agreement between RDMs is interpreted against a noise ceiling. Cross-validated estimators such as crossnobis reduce the bias that noise introduces in the measurement \citep{walther2016,nili2014}. This work applies the same principles to answer the alignment question.

\section{Methods}
\label{sec:measurements}

\paragraph{Datasets and representation.}
Three motor-imagery cohorts are analyzed: an in-house dry-electrode dataset \citep{mortensen2026}, Weibo2014 \citep{yi2014}, and BCI Competition IV-2a \citep{brunner2008}, the latter two accessed through MOABB \citep{jayaram2018, aristimunha2023moabb}. The cohorts were recorded with different montages, so the analysis retains a common subset of eight electrodes present in all three. Each trial is summarized by a regularized spatial covariance \citep{ledoit2004,chen2010}, so every trial lies on $\mathcal{P}_8$ and all three cohorts can be analyzed within one geometry. \appendixref{app:data} gives the full dataset and preprocessing details.

\paragraph{Alignment operations.}
For each subject $s$, per-subject recentering is based on the subject's own
Fr\'{e}chet mean $\mathbf{\bar C}_s$, defining the recentering matrix
$\mathbf{W}_s=\mathbf{\bar C}_s^{-1/2}$, which maps a covariance matrix $\mathbf{C}$ to
$\mathbf{W}_s \mathbf{C W}_s^{\top}$, sending the subject's mean covariance to the identity
\citep{zanini2018}. Global recentering applies the same operation to every
subject and therefore leaves the displacement between subject means in place,
serving as control condition throughout the analysis. Comparing the two recentering approaches isolates the effect of subject-specific alignment.

\paragraph{Decoding models and evaluation.}
Four covariance-based decoders are compared: Minimum Distance to Riemannian
Mean (MDM) and tangent-space logistic regression, both fixed-metric baselines
\citep{barachant2012}; SPDNet, which learns bilinear maps between SPD matrices
\citep{huang2017}; and TSMNet, which learns a per-domain recentering inside the
network \citep{kobler2022}. EA+EEGNet, a convolutional decoder applied after
Euclidean alignment, is included as a non-covariance reference
\citep{he2020,lawhern2018}. All models are evaluated with leave-one-subject-out
(LOSO) balanced accuracy, computed for each subject and then averaged across
subjects; stochastic models use three seeds and every comparison uses the same
folds and scoring unit (\appendixref{app:decoding}).

\paragraph{Primary relational measure -- AIRM condition RDM.}
The primary analysis constructs the RDM directly from the AIRM distance, so that the invariance result of \sectionref{sec:theory} applies to it exactly. After global or per-subject recentering, the five window covariances from each trial are combined with an AIRM Fr\'{e}chet mean. The resulting trial covariances are then averaged within each subject and condition to obtain $\mathbf{P}_{s,c}\in\mathcal{P}_d$, \emph{condition mean} for subject $s$ and condition $c$. Both Fréchet mean aggregation steps are equivariant under congruence, and therefore applying a congruence before aggregation is equivalent to applying the same congruence to the resulting trial and condition means. \appendixref{app:proofs} gives the argument. The RDMs use condition means rather than single trials because single-trial clusters overlap strongly in this regime (\appendixref{app:separability}). The RDM for subject $s$ is
\begin{equation}
\mathbf{D}_s(i,j)=\delta\bigl(\mathbf{P}_{s,i},\mathbf{P}_{s,j}\bigr),
\label{eq:rdm}
\end{equation}
where $i$ and $j$ index task conditions, $\mathbf{P}_{s,i}$ and $\mathbf{P}_{s,j}$ are that subject's
condition means for conditions $i$ and $j$, and $\delta$ is the AIRM distance of
\sectionref{sec:background}.
Cross-subject agreement $\rho$ is the mean pairwise Spearman correlation between subjects' RDMs. The lower noise ceiling $\kappa$ correlates each RDM with the mean RDM of the remaining subjects, providing an estimator-specific measure of group RDM reliability.
The agreement-to-ceiling difference is defined as $g=\kappa-\rho$ and is reported as a descriptive diagnostic. Because $\rho$ compares individual subject RDMs whereas $\kappa$ compares each subject with an average over the remaining subjects, the two
quantities have different noise properties. Consequently, $g$ is not interpreted as a direct
estimate of latent subject-specific relational structure.

\paragraph{Complementary measure -- crossnobis.}
A second analysis constructs RDMs from trial-level tangent vectors with cross-validated Mahalanobis distance \citep{walther2016,nili2014}; Ledoit-Wolf
shrinkage provides the noise precision. The AIRM theorem does not imply exact invariance for this estimator, so crossnobis constitutes an independent, noise-normalized robustness test. A separate subject probe measures how much subject identity is retained in each representation.

\paragraph{Training for cross-subject similarity.}
Rank-based RSA and the bootstrapped difference are not convenient differentiable objectives, so CKA is
used as a differentiable surrogate (dCKA) for cross-subject agreement \citep{kornblith2019}. Adding dCKA to the loss makes the network optimize directly for cross-subject similarity during training, rather than receiving it only as a by-product of classification. The training loss is
\begin{equation}
\mathcal{L} = \mathcal{L}_{\mathrm{CE}} + \lambda\,\mathcal{L}_{\mathrm{dCKA}},
\label{eq:loss}
\end{equation}
where $\mathcal{L}_{\mathrm{CE}}$ is the cross-entropy classification loss and
$\mathcal{L}_{\mathrm{dCKA}} = 1-\mathrm{CKA}$, so that minimizing the loss
increases cross-subject CKA. The sweep uses $\lambda\in\{0,1,10,30\}$, three seeds, and nineteen LOSO folds. The $\lambda=0$ runs follow the same code path, and all analyses use held-out learned features
(\appendixref{app:dcka}).

\section{Results}
\label{sec:results}

\subsection{Decoding performance after alignment}
\label{sec:saturation}

\tableref{tab:saturation} reports aligned LOSO balanced accuracy. The four covariance
methods differ by at most 0.017 on either gel cohort and 0.036 in-house. These margins are
small but not negligible: three internal contrasts remain significant
(\appendixref{app:decoding}), so we do not claim that the methods are equivalent. Alignment
nonetheless has a larger and more consistent effect than the choice of decoder, which
motivates analyzing the representation geometry directly.
\begin{table}[htbp]
\centering
\caption{
Aligned LOSO balanced accuracy is reported. For stochastic models, results are averaged over three random seeds for each subject. $N$ denotes the dataset size. Complete statistical analyses and unaligned results are provided in \appendixref{app:decoding}.}
\label{tab:saturation}
\small
\setlength{\tabcolsep}{4pt}
\begin{tabular}{@{}lccc@{}}
\toprule
 & BCIC 4-class & Weibo 4-class & In-house 5-class \\
 & $N{=}9$, chance $0.25$ & $N{=}10$, chance $0.25$ & $N{=}19$, chance $0.20$ \\
\midrule
MDM & 0.453 & 0.425 & 0.242 \\
Tangent space + LR & 0.456 & 0.442 & 0.278 \\
SPDNet & 0.455 & 0.440 & 0.245 \\
TSMNet (SPDDSMBN) & 0.462 & 0.439 & 0.258 \\
\midrule
EA + EEGNet & 0.412 & 0.382 & 0.288 \\
\bottomrule
\end{tabular}
\vspace{-0.6cm}
\end{table}

\subsection{The dissociation}
\label{sec:dissociation}
\begin{figure}[!t]
\centering
\includegraphics[width=\linewidth]{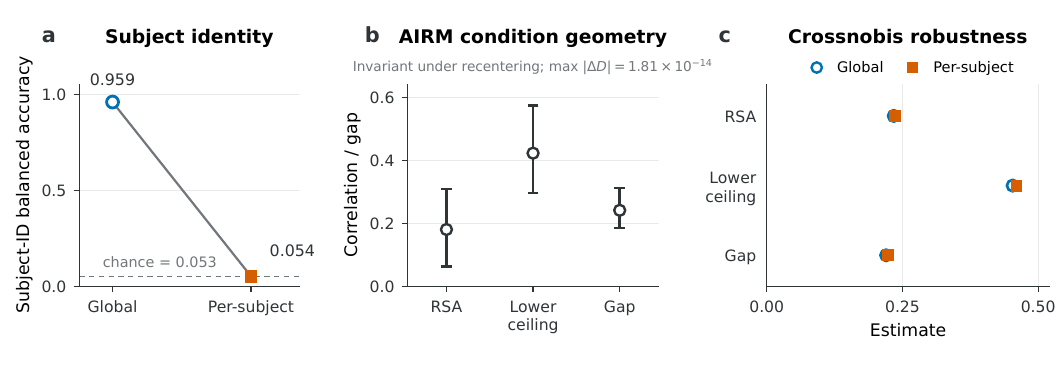}
\caption{Per-subject alignment removes subject identity without altering the
theorem-matched relational geometry. (a) The subject-identification probe collapses
to chance. (b) The AIRM RDM is invariant under recentering; the largest entrywise
change is $1.81\times10^{-14}$. (c) Crossnobis behaves the same way, though the
theorem does not cover it. Values and intervals are given in the text.}
\vspace{-0.9cm}
\label{fig:fig1}
\end{figure}

The identity axis changes sharply. The subject probe falls from 0.959 to 0.054, essentially reaching its chance level of 0.053 (\figureref{fig:fig1}a). Therefore, single trials cease to identify their source subjects, and TSMNet tangent features show the same collapse (\appendixref{app:estimability}). The identity result replicates on both public cohorts: the probe changes from 0.978 to 0.076 on BCIC, with chance at 0.111, and from 0.981 to 0.066 on Weibo, with chance at 0.100.

The relational axis behaves differently. For the AIRM condition RDM, agreement is $\rho=0.181$ with 95\% CI $[0.064,0.310]$ and permutation $p<0.001$. The lower ceiling is $\kappa=0.423$ with 95\% CI $[0.297,0.573]$, which leaves a difference of $g=0.242$ with 95\% CI $[0.185,0.313]$ (\figureref{fig:fig1}b). Global and per-subject recentering produce RDMs that differ by at most $1.81\times10^{-14}$, so $\rho$, $\kappa$, and $g$ are identical at the reported precision. Therefore, subjects show significant cross-subject agreement in condition geometry, since $\rho$ is significantly above zero, while the observed agreement remains below the estimator-specific lower noise ceiling. The difference between these quantities is reported
descriptively and is not interpreted as the fraction of reliable structure that is subject specific.

Crossnobis provides an independent check outside the theorem-matched metric. Agreement changes only from 0.234 to 0.236, while $g$ changes from 0.220 to 0.224 (\figureref{fig:fig1}c); both shifts lie within estimation noise.

\subsection{Congruence invariance of the AIRM RDM}
\label{sec:theory}

The numerical equality above is not a coincidence. It can be derived using the definitions of \sectionref{sec:measurements}, from the geometry of the metric itself.
\vspace{-0.2cm}
\begin{definition}
A \emph{per-subject congruence} is a family of maps
$\alpha_s(\mathbf{P})=\mathbf{E}_s\mathbf{PE}_s^{\top}$, with one invertible matrix
$\mathbf{E}_s\in\mathrm{GL}(d)$ for each subject, where $\mathrm{GL}(d)$ denotes the set of invertible $d\times d$ matrices.
\end{definition}
\vspace{-0.2cm}

Per-subject recentering belongs to this class with $\mathbf{E}_s=\mathbf{\bar C}_s^{-1/2}$, as does geodesic transport to a common basepoint. Orthogonal conjugation in the tangent space, $\mathbf{S}\mapsto \mathbf{R}_s\mathbf{SR}_s^{\top}$, also has this form, whereas an arbitrary rotation of the vectorized tangent coordinates does not.
\vspace{-0.2cm}
\begin{theorem}[Congruence invariance of the AIRM RDM]
\label{thm:invariance}
Let $\{\alpha_s\}$ be per-subject congruences, and let $\mathbf{D}'_s$ use $\alpha_s(\mathbf{P}_{s,c})$ in place of $\mathbf{P}_{s,c}$. Then $\mathbf{D}'_s=\mathbf{D}_s$ for every subject. Consequently, $\rho$, $\kappa$, and $g$ are unchanged.
\end{theorem}
\vspace{-0.2cm}

The proof is based on the isometry property $\delta(\mathbf{EPE}^{\top},\mathbf{EQE}^{\top})=\delta(\mathbf{P},\mathbf{Q})$ of the AIRM distance \citep{congedo2017}, and is given in \appendixref{app:proofs}. The property itself is known: recentering preserves within-subject AIRM distances \citep{zanini2018,he2020}, and \citet{he2020} note that Euclidean alignment can leave per-class distributions mismatched across subjects, consistent with the present result. Its consequence for representation measurement is central: because Fr\'{e}chet-mean aggregation is equivariant under congruence, the invariance carries over to the condition RDM, which accounts exactly for the numerical equality reported above. It does not cover crossnobis, whose similar behavior remains an empirical robustness result.

The same argument applies to shared congruence maps, obtained by setting $\mathbf{E}_s=\mathbf{F}$ for every subject; shared whitening, an SPD bias, or a square orthogonal BiMap acting as a congruence also leave the AIRM RDM unchanged. It is worth noting that this derives from the congruence structure of these transformations, not from their being shared across subjects; more general shared maps need not preserve the RDM. A shared \emph{dispersion} $\phi_{\gamma}(\mathbf{P})=\mathbf{P}^{\gamma}$, which contracts or expands the representation around the identity, is one such map: it is not a congruence and can change AIRM distances. On the other hand, it preserves the rank structure of the RDM, and hence $\rho$, $\kappa$, and $g$, whenever each subject's condition means commute (\lemmaref{lem:dispersion}, \appendixref{app:proofs}).

\subsection{Training for cross-subject similarity does not increase held-out agreement}
\label{sec:training}

The identity probe remains at chance in all twelve runs. As $\lambda$ increases from 0 to 30, the mean difference falls by 0.057 (\figureref{fig:fig2} in \appendixref{app:dcka}), but mean RSA also falls from 0.153 to 0.135 while the lower ceiling drops by 0.074. Across runs, $g$ correlates strongly with the lower ceiling ($r=0.888$). The smaller observed difference therefore results from the falling ceiling rather than from improved held-out agreement. A ceiling-normalized ratio has the same failure mode whenever the ceiling falls faster than agreement. The same statistic fails in the opposite direction when agreement is inflated rather than reliability reduced, which \appendixref{app:ann} demonstrates against known ground truth.

Training increases batch-level CKA, confirming that the surrogate receives a usable gradient. However, increasing its weight thirtyfold leaves the final in-sample CKA near 0.545, while balanced accuracy changes only from 0.248 to 0.236. This pattern is consistent with \citet{davari2023reliability}, who show that CKA can change substantially without a corresponding change in network function. Optimizing the surrogate can therefore raise in-batch CKA without
altering the held-out relational structure that the RSA measures. These results argue against the explanation that the similarity objective was simply too weakly weighted.

The learned-feature estimates also vary substantially across seeds. At $\lambda=0$, the lower ceiling ranges from 0.252 to 0.456, and a crossed analysis attributes 57--69\% of the variance in the representational measures to the seed (\appendixref{app:dcka}). Seed averages are therefore emphasized, and \figureref{fig:fig2} shows the individual trajectories.

\subsection{Sharedness alone does not explain the result}
\label{sec:sharedness}

The transformations tested above are subject-specific or shared, and none increases
cross-subject relational agreement. This does not imply that sharedness is what protects
the RDM.
\vspace{-0.3cm}
\begin{proposition}[A shared finite-sample map]
\label{prop:counterexample}
Some subject condition configurations cannot be related by any per-subject congruence
because their AIRM distance structures differ. Nevertheless, there exists a smooth shared
map that sends their finite sets of condition means to the same target configuration. The
resulting RDMs are identical, so cross-subject agreement is maximal.
\end{proposition}

\appendixref{app:proofs} constructs such a map in the matrix-logarithm space. Since a transformation with
no subject-specific parameters can still drive agreement to its maximum, sharedness is not
the reason the tested maps leave the RDM unchanged. The construction also marks the
boundary of the present result: it need only interpolate a finite set of condition means,
whereas a useful transformation must generalize to unseen trials while preserving reliable
task-related structure. Whether restricted shared transformations can increase held-out
agreement under that requirement remains open.
\section{Discussion and limitations}
\label{sec:discussion}

\paragraph{Effect of alignment.}
Per-subject alignment moves subject distributions toward a common reference and removes
subject identity. For cross-subject decoding, it therefore removes the distributional
offsets that make trials subject-identifiable and improves decoding in the tested settings,
while leaving the AIRM condition RDM unchanged. Explicit CKA-based training likewise does
not increase held-out relational agreement. Distributional de-identification and relational alignment are distinct properties,
as their definitions suggest; the present experiments quantify the distinction but do not determine whether a more expressive subject-adaptive map can improve both simultaneously.

\paragraph{Scope of the alignment class.}
\theoremref{thm:invariance} covers every per-subject congruence, including Euclidean alignment and the rotation step of Riemannian Procrustes analysis (RPA) \citep{rodrigues2019}, since an orthogonal conjugation is a congruence. Such maps change where each subject's condition configuration sits relative to the others, which a shared decoder can exploit, but not the configuration itself, which is what the RDM measures. The stretching step of RPA is a per-subject power map outside this class; \lemmaref{lem:dispersion} covers it only for a shared exponent and commuting condition means, so its effect on relational agreement remains open.

\paragraph{Cohort differences and scope.}
The identity collapse replicates across all three cohorts, but the agreement-to-ceiling difference can be
estimated reliably only on the in-house cohort. Reliable RDM estimation requires both enough
conditions, which set the number of informative RDM entries, and enough subjects, which set
the number of RDM pairs contributing to the group estimate
(\appendixref{app:estimability}). At a comparable channel count, the available public
cohorts do not provide both simultaneously, whereas the identity axis needs neither and
behaves identically everywhere. This is therefore a property of the public datasets rather
than of the effect, and replicating the relational claim requires new data with more
conditions per subject.

\paragraph{Implication for representation studies.}
The isometry argument extends beyond EEG, but the invariance class that matters belongs to the whole measurement, not to the metric alone. The AIRM makes each RDM invariant to per-subject congruences, and the rank-based agreement $\rho$ is further unchanged by any per-subject transform that preserves the order of a subject's RDM entries; the dispersion map of \lemmaref{lem:dispersion} is a non-isometric example. An alignment claim must therefore be matched to the invariance class of both the distance and the statistic. Agreement, ceiling, and their difference should also be reported together, because a smaller observed difference can result either from increasing agreement or from a falling reliability ceiling. Both failure modes are demonstrated against known ground truth in
\appendixref{app:ann}: an isometric alignment leaves agreement fixed, and a sufficiently
flexible one drives it above the structure the domains actually share.

\paragraph{Limitations.} The relational estimate is reliable only on one cohort. The dCKA result is also narrow in scope: it covers one architecture (TSMNet
with scale-only SPDDSMBN), the in-house cohort, and three seeds, and seed effects explain 57--69\% of the variance in the representational measures
(Appendix~\ref{app:dcka}). Theorem~\ref{thm:invariance} applies exactly to the
AIRM condition RDM but not to crossnobis, dimension-reducing BiMaps, or optimization with a relational surrogate, so the dCKA result is empirical
rather than a general impossibility result. Finally, $g$ is not calibrated against its expected value under identical true RDMs, which is positive because $\rho$ and $\kappa$ have different noise properties, so its magnitude is not interpreted.

\section{Conclusion and future work}
\label{sec:conclusion}

Cross-subject alignment can remove subject-identifying structure without changing relational
geometry measured by an invariant metric. On the in-house cohort, recentering leaves the
theorem-matched AIRM RDM unchanged to numerical precision, while cross-subject
agreement remains significantly above zero but below the estimator-specific lower noise
ceiling. Explicit relational pressure through a differentiable CKA surrogate also does not
increase held-out agreement in the tested model. These findings separate distributional
de-identification from relational alignment and show that an alignment claim must be interpreted relative to the
invariances of both the metric and the statistic used to evaluate it. Future work should test whether the dissociation persists at higher channel counts and accuracies, evaluate alignment when a small labeled calibration set from the new subject is available, and extend the invariance analysis to estimators such as crossnobis.


\newpage
\acks{This work was carried out during a research internship at the Department
of Health Technology, Technical University of Denmark (DTU). 
A.N. is supported by a scholarship from the Konrad Zuse School of
Excellence in Learning and Intelligent Systems (ELIZA), as part of the funding
programme Konrad Zuse Schools of Excellence in Artificial Intelligence, financed
by the German Academic Exchange Service (DAAD) and the German Federal Ministry of
Research, Technology and Space (BMFTR).}

\appendix
\newpage

\section{Datasets and Preprocessing}
\label{app:data}

\paragraph{BCI Competition IV-2a.}
This public cohort contains data from nine subjects recorded with 22 gel electrodes, four
MI classes, and 4\,s trials \citep{brunner2008}. The data are
available through MOABB \citep{jayaram2018}. Of the 22 available electrodes,
only those matching the common eight-channel montage were retained: Fz, Cz, C3,
C4, CP1, and CP2 were used directly, while T7 and T8, which this cohort does not
record, were replaced by the nearby electrodes C5 and C6.

\paragraph{Weibo2014.}
This public cohort contains ten subjects recorded with gel electrodes and 4\,s
trials \citep{yi2014}. It includes both simple and compound MI
tasks, together with a rest condition. Rest is excluded from decoding, which
uses the four simple conditions. The relational analysis instead restores the
two compound-limb tasks, giving six distinct motor conditions: left hand, right
hand, both hands, feet, left hand with right foot, and right hand with left
foot. The larger number of conditions yields a larger RDM and therefore a more
stable relational estimate.

\paragraph{In-house cohort.}
The dry-electrode cohort contains nineteen subjects sampled at 250\,Hz with
5\,s trials. Seven classes were recorded: right hand, left hand, both hands,
feet, and tongue MI, together with rest and an imagined-blink class.
The relational analysis uses the five MI conditions only. Rest and
imagined blink were excluded because including these classes
lowered the lower noise ceiling to the point where cross-subject agreement was
no longer reliably estimable. The in-house cohort was recorded at DTU Health Tech as part of a bachelor thesis
\citep{mortensen2026}, using an OpenBCI Ultracortex Mark IV headset with
ThinkPulse active dry electrodes and a Cyton board. All participants were healthy
volunteers who gave written informed consent; data were processed in accordance
with the GDPR and stored pseudonymized.

\paragraph{Common montage.}
All cohorts use Fz, Cz, C3, C4, T7, T8, CP1, and CP2, which yields a common eight-channel covariance representation.

\paragraph{Preprocessing.}
Preprocessing of the EEG data includes channel selection, a 6--32\,Hz FIR
bandpass filter, a 50\,Hz notch filter, resampling to 250\,Hz, and channel
demeaning. The pipeline deliberately omits Euclidean alignment, channel
z-scoring, and common average reference, since each conflicts with the
properties the covariance representation relies on: Euclidean alignment is
itself a per-subject alignment and would confound the study target, channel
z-scoring applies a diagonal per-domain transform, and common average reference
makes every covariance rank deficient. Amplitude-based trial rejection is also
disabled, because its thresholds are defined in standardized units that the
pipeline does not produce. The preprocessing therefore differs from that of the
related work in \sectionref{sec:background}, where these steps are standard. 

The pipeline uses 2\,s windows with 75\% overlap, giving 500 samples per window.
Each window covariance is estimated with the oracle approximating shrinkage
(OAS) estimator \citep{chen2010}, a shrinkage method that regularizes the
sample covariance toward a scaled identity matrix and so guarantees a
well-conditioned SPD result from short windows. Ledoit-Wolf shrinkage
\citep{ledoit2004} separately estimates the noise precision used by the
crossnobis distance.

\paragraph{AIRM input.}
The full in-house SPD set contains $33{,}950$ window covariances from $6{,}790$
trials, with five windows per trial. The main AIRM analysis retains the five
MI classes and removes rest and the imagined-blink class. The
resulting set contains $24{,}250$ windows from $4{,}850$ trials, and every
covariance is SPD.

\newpage
\section{Decoding protocol and full comparisons}
\label{app:decoding}
\paragraph{Protocol.}
Each model is trained on 2\,s windows. Within a trial, the window probabilities
are averaged and the class with the highest mean probability is selected.
Balanced accuracy is computed for each subject and then averaged across
subjects. Stochastic models use three seeds, and every comparison uses the same
folds and scoring unit.

The convolutional reference follows the inherited EA+EEGNet pipeline
\citep{he2020,lawhern2018}. Reproducing its binary task gives
$0.669\pm0.111$ against a published $0.679$, so the implementation behaves as
expected and the small shortfall is consistent with library-version differences.
This reference was tuned by its authors for binary decoding and was not retuned
for the four- and five-class tasks studied here. Any margin the covariance
methods show over it is therefore a margin over an untuned baseline, and should
not be read as evidence that covariance methods outperform a well-tuned
convolutional network.

The inherited Euclidean alignment computes one covariance from each continuous run, whereas the canonical method averages trial covariances \citep{he2020}. The inherited variant is retained for reproducibility.

\paragraph{Effect of alignment.}
Per-subject recentering improves all eight fixed-metric contrasts. On BCIC, MDM rises from 0.318 to 0.453 ($p=0.004$), while the tangent-space method rises from 0.356 to 0.456 ($p=0.008$). On Weibo, the corresponding changes are 0.354 to 0.425 ($p=0.014$) and 0.368 to 0.442 ($p=0.004$).

The in-house gains are smaller: tangent-space accuracy increases by 0.030 ($p=0.031$), whereas MDM increases by 0.020 ($p=0.219$). This pattern suggests that a first-order mean offset does not explain most of the in-house variation between subjects. Recentering changes decoding accuracy while leaving the AIRM condition RDM unchanged because the two depend on different geometry: a LOSO decoder applies a single classifier to all subjects and is therefore sensitive to where each subject's covariances lie relative to the others, whereas the RDM
depends only on distances among a subject's own condition means (Theorem~\ref{thm:invariance}). No tested method closes the
agreement-to-ceiling difference $g$; explicit relational training is
examined in Section~\ref{sec:training}.

\paragraph{Aligned-method tests.}
No contrast among aligned methods exceeds 0.017 on BCIC or Weibo, and none is significant. Three in-house contrasts are significant: SPDNet is 0.033 below tangent space ($p=0.001$), TSMNet is 0.020 below tangent space ($p=0.004$), and TSMNet is 0.013 above SPDNet ($p=0.025$).

A full-rank control checks the SPDNet implementation. Full-rank SPDNet matches tangent-space accuracy within a few thousandths on every cohort because LogEig commutes with orthogonal conjugation and the linear head absorbs the resulting reparameterization. The observed deficits arise from dimension reduction: reducing $8\times8$ matrices to $4\times4$ costs 0.035 in-house ($p=0.010$), while reducing to $2\times2$ costs 0.072 on Weibo ($p=0.002$).

\paragraph{Convolutional reference.}
The recentered tangent-space method exceeds EA+EEGNet by 0.044 on BCIC ($p=0.027$) and by 0.059 on Weibo ($p=0.037$), while the in-house difference is $-0.010$ ($p=0.145$). MDM follows the same direction on the gel cohorts without reaching significance, and it is 0.046 below EEGNet in-house ($p=0.001$).

A mixed model pools the three multiclass settings on a chance-normalized scale. The tangent-space effect is 0.029 with 95\% CI $[0.003,0.054]$ and $p=0.029$, whereas the MDM effect is 0.000 with $p=0.980$. Because the model includes only three dataset groups, it should be read as a summary of the paired results rather than as an independent test. Relative to EEGNet, TSMNet differs by 0.050 on BCIC, 0.056 on Weibo, and $-0.030$ in-house.

\newpage
\section{Cross-cohort identity collapse and relational estimability}
\label{app:estimability}

\begin{table}[h]
\centering
\caption{Theorem-matched AIRM RDM results. Raw, global, and per-subject representations give identical values at the reported precision. Intervals are subject-bootstrap 95\% confidence intervals.}
\label{tab:airm}
\begin{tabular}{lcc}
\toprule
Quantity & Estimate & 95\% CI \\
\midrule
Cross-subject RSA $\rho$ & 0.181 & $[0.064, 0.310]$ \\
Lower noise ceiling $\kappa$ & 0.423 & $[0.297, 0.573]$ \\
Upper noise ceiling & 0.463 & $[0.381, 0.596]$ \\
Agreement-to-ceiling difference $g=\kappa-\rho$ & 0.242 & $[0.185, 0.313]$ \\
Permutation $p$ (RSA) & \multicolumn{2}{c}{$1.9996 \times 10^{-4}$} \\
\midrule
\multicolumn{3}{l}{\emph{Congruence diagnostics for entrywise RDM differences}} \\
raw vs. global, max $|\Delta \mathbf{D}|$ & \multicolumn{2}{c}{$2.85 \times 10^{-14}$} \\
raw vs. per-subject, max $|\Delta \mathbf{D}|$ & \multicolumn{2}{c}{$3.64 \times 10^{-14}$} \\
global vs. per-subject, max $|\Delta \mathbf{D}|$ & \multicolumn{2}{c}{$1.81 \times 10^{-14}$} \\
$\Delta\rho$, $\Delta\kappa$, $\Delta g$ & \multicolumn{2}{c}{$0$ at reported precision} \\
\bottomrule
\end{tabular}
\end{table}

\begin{table}[h]
\centering
\caption{Complementary crossnobis results. \theoremref{thm:invariance} does not cover this estimator, and the observed changes remain within estimation noise.}
\label{tab:crossnobis}
\begin{tabular}{lccc}
\toprule
 & Global & per-subject & Change \\
\midrule
RSA $\rho$ & 0.234 & 0.236 & $+0.002$ \\
Lower ceiling $\kappa$ & 0.453 & 0.460 & $+0.007$ \\
$g$ & 0.220 $[0.157, 0.299]$ & 0.224 $[0.156, 0.298]$ & $+0.004$ \\
\bottomrule
\end{tabular}
\end{table}

\begin{table}[htb!]
\centering
\caption{Subject-identity probe across cohorts. Per-subject recentering reduces the probe to chance in every cohort.}
\label{tab:probe}
\begin{tabular}{lccc}
\toprule
Cohort ($N$, conditions) & Global & per-subject & Chance \\
\midrule
In-house (19, 5) & 0.959 & 0.054 & 0.053 \\
BCIC (9, 4) & 0.978 & 0.076 & 0.111 \\
Weibo (10, 6) & 0.981 & 0.066 & 0.100 \\
\bottomrule
\end{tabular}
\end{table}

\newpage

\begin{table}[htb!]
\centering
\caption{Crossnobis estimability across cohorts. A negative or small lower ceiling indicates that the relational quantity is poorly estimated.}
\label{tab:estimability}
\begin{tabular}{lccccc}
\toprule
Cohort & $N$ & $C$ & RSA [95\% CI] & Lower ceiling & Perm. $p$ \\
\midrule
Weibo (4 cond.) & 10 & 4 & 0.011 $[-0.190, 0.272]$ & $-0.051$ & 0.356 \\
Weibo (6 cond.) & 10 & 6 & 0.066 $[-0.058, 0.235]$ & 0.156 & 0.071 \\
BCIC & 9 & 4 & 0.202 $[-0.096, 0.482]$ & 0.467 & 0.013 \\
In-house & 19 & 5 & 0.236 $[0.080, 0.398]$ & 0.460 & $2\times10^{-4}$ \\
\bottomrule
\end{tabular}
\end{table}

Writing $C$ for the number of task conditions, each RDM contains $C(C-1)/2$ informative off-diagonal entries, while the subject count determines how many RDM pairs contribute to the group estimate. Stable estimation requires both quantities to be large enough. At a comparable channel count, the available public datasets do not provide enough subjects and motor conditions simultaneously.

\newpage
\section{Proofs}
\label{app:proofs}
\setcounter{equation}{0}
\renewcommand{\theequation}{D\arabic{equation}}

\paragraph{Congruence equivariance of the AIRM Fréchet mean.}
Let
\begin{equation}
\mathbf{M}=\operatorname*{arg\,min}_{\mathbf{X}\in\mathcal{P}_d}
\sum_i \delta^2(\mathbf{X},\mathbf{P}_i)
\end{equation}
be the AIRM Fréchet mean of $\{\mathbf{P}_i\}$, with $\mathbf{X} \in \mathcal{P}_d$ as optimization variable. For any invertible matrix $\mathbf{E}$,
\begin{equation}
\operatorname{FM}_{\mathrm{AIRM}}\{\mathbf{EP}_i\mathbf{E}^\top\}=\mathbf{EME}^\top.
\end{equation}
Indeed, congruence invariance of the AIRM gives
\begin{equation}
\sum_i \delta^2(\mathbf{X},\mathbf{EP}_i\mathbf{E}^\top)=\sum_i \delta^2(\mathbf{E^{-1}XE}^{-\top},\mathbf{P}_i).
\end{equation}
The minimizers therefore correspond under
$\mathbf{X}\mapsto \mathbf{E^{-1}XE}^{-\top}$. Consequently, both the window to trial and
trial to condition Fréchet means used in the empirical pipeline transform by the same
congruence.

\paragraph{Proof of \theoremref{thm:invariance}.}
The AIRM distance satisfies $\delta(\mathbf{EPE}^{\top},\mathbf{EQE}^{\top})=\delta(\mathbf{P},\mathbf{Q})$. Applying this identity with $\mathbf{E}=\mathbf{E}_s$ to each condition pair gives\\ $[\mathbf{D}'_s]_{ij}=\delta(\mathbf{E}_s\mathbf{P}_{s,i}\mathbf{E}_s^{\top},\mathbf{E}_s\mathbf{P}_{s,j}\mathbf{E}_s^{\top})=[\mathbf{D}_s]_{ij}$, so $\mathbf{D}'_s=\mathbf{D}_s$. Since $\rho$ and $\kappa$ depend only on the RDMs, $g=\kappa-\rho$ is unchanged as well. \hfill$\blacksquare$

\begin{lemma}[Commuting dispersion]
\label{lem:dispersion}
Assume that each subject has Fr\'{e}chet mean $\mathbf{I}$, and let a shared $\gamma>0$ define $\phi_{\gamma}(\mathbf{P})=\mathbf{P}^{\gamma}$. If one subject's condition means commute, then $\mathbf{D}^{\phi_{\gamma}}_s=\gamma \mathbf{D}_s$. If this condition holds for every subject, then $\rho$, $\kappa$, and $g$ remain unchanged. The condition is restrictive: condition mean covariances do not generally commute in EEG.
\end{lemma}

\paragraph{Proof of \lemmaref{lem:dispersion}.}
Fix one subject and two condition means $\mathbf{A}$ and $\mathbf{B}$. If they commute, a single orthogonal matrix diagonalizes both, so $\mathbf{A}=\mathbf{U\Lambda U}^{\top}$ and $\mathbf{B}=\mathbf{UMU}^{\top}$. It follows that $\mathbf{A}^{-\gamma/2}\mathbf{B}^{\gamma}\mathbf{A}^{-\gamma/2}=\mathbf{U}(\mathbf{\Lambda}^{-1}\mathbf{M})^{\gamma}\mathbf{U}^{\top}$, whose principal logarithm is $\gamma \mathbf{U}\log(\mathbf{\Lambda}^{-1}\mathbf{M})\mathbf{U}^{\top}$. Therefore $\delta(\mathbf{A}^{\gamma},\mathbf{B}^{\gamma})=\gamma\delta(\mathbf{A},\mathbf{B})$.

Under mutual commutation, this identity holds for every condition pair and gives $\mathbf{D}_s^{\phi_{\gamma}}=\gamma \mathbf{D}_s$. Spearman correlation is invariant to positive scaling, so $\rho$ remains constant. The shared scale also multiplies the group RDM and therefore leaves $\kappa$, and hence $g$, unchanged. \hfill$\blacksquare$

\paragraph{Proof of Proposition~\ref{prop:counterexample}.}
Consider two subjects with different AIRM distance matrices. No congruence relates their configurations because every congruence preserves all AIRM distances. Their $2C$ condition means form a finite set of distinct points. In the global logarithm chart from $\mathcal{P}_d$ to $\mathrm{Sym}(d)$, choose one target $\mathbf{Q}_c$ for each condition and construct a smooth finite interpolation that maps both subject points for condition $c$ to $\log \mathbf{Q}_c$.

With $\phi=\exp\circ\psi\circ\log$, both subjects satisfy $\phi(\mathbf{P_{1,c}})=\phi(\mathbf{P}_{2,c})=\mathbf{Q}_c$ for every condition. Their post-map RDMs are therefore identical, which gives $\rho=1$, $\kappa=1$, and $g=0$. \hfill$\blacksquare$

\newpage
\section{dCKA details}
\label{app:dcka}

\begin{figure}[htbp]
\centering
\includegraphics[width=0.95\linewidth]{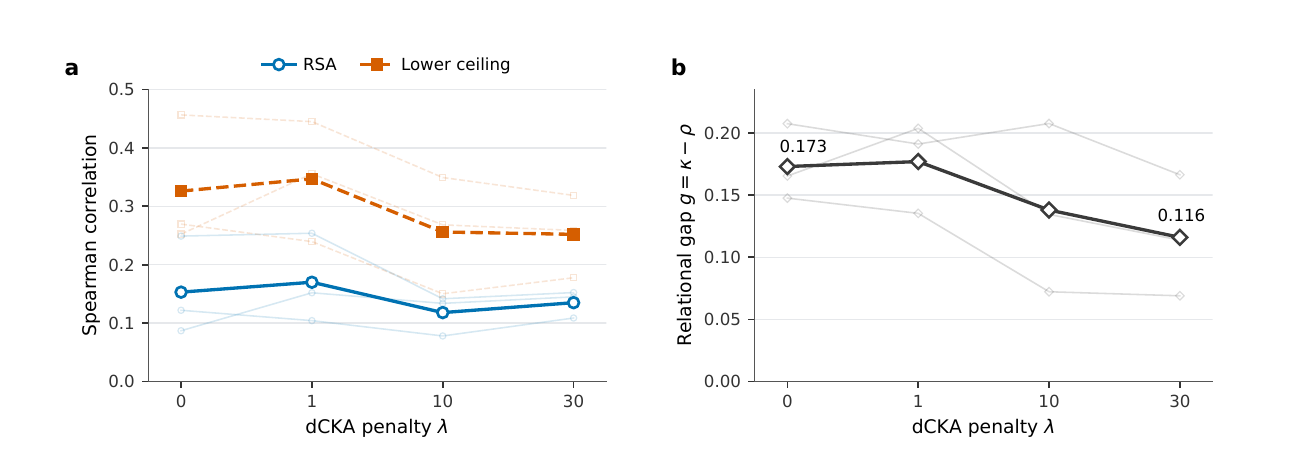}
\caption{Effect of the dCKA surrogate across three seeds. From $\lambda=0$ to 30, mean RSA changes from 0.153 to 0.135, while the mean lower ceiling falls from 0.326 to 0.252. The mean difference therefore decreases from 0.173 to 0.116. Faint trajectories show the substantial variation across seeds.}
\vspace{-0.6cm}
\label{fig:fig2}
\end{figure}

\begin{table}[h]
\centering
\caption{dCKA sweep. Each entry is the mean and standard deviation across three seeds; probe chance is 0.053.}
\label{tab:dcka}
\begin{tabular}{cccccc}
\toprule
$\lambda$ & BAcc & RSA $\rho$ & Lower ceiling $\kappa$ & $g$ & Probe \\
\midrule
0  & $0.248 \pm 0.015$ & $0.153 \pm 0.085$ & $0.326 \pm 0.113$ & $0.173 \pm 0.031$ & 0.052 \\
1  & $0.239 \pm 0.009$ & $0.170 \pm 0.076$ & $0.347 \pm 0.103$ & $0.177 \pm 0.036$ & 0.053 \\
10 & $0.234 \pm 0.004$ & $0.118 \pm 0.035$ & $0.256 \pm 0.100$ & $0.138 \pm 0.068$ & 0.052 \\
30 & $0.236 \pm 0.005$ & $0.135 \pm 0.023$ & $0.252 \pm 0.071$ & $0.116 \pm 0.049$ & 0.053 \\
\bottomrule
\end{tabular}
\end{table}

\begin{table}[h]
\centering
\caption{In-sample CKA. Training increases CKA, but the final value remains almost unchanged as the penalty weight increases.}
\label{tab:insample}
\begin{tabular}{ccc}
\toprule
$\lambda$ & In-sample $\Delta$CKA & Maximum CKA \\
\midrule
1  & $+0.035 \pm 0.004$ & 0.541 \\
10 & $+0.040 \pm 0.001$ & 0.545 \\
30 & $+0.039 \pm 0.002$ & 0.544 \\
\bottomrule
\end{tabular}
\end{table}

All runs use scale-only SPDDSMBN and the same code path, with $\lambda=0$ as the on-path control. Across the twelve runs, $g$ correlates with the lower ceiling at $r=0.888$. Seed effects explain 57--69\% of the variance in the representational measures, whereas penalty effects explain 13--29\%. At $\lambda=0$, the lower ceiling ranges from 0.252 to 0.456.

The main text reports seed averages. A single TSMNet run gives an AIRM-side $g$ of 0.246, which lies within the seed-averaged interval for $\lambda=0$.

\newpage
\section{Separability of the per-trial representation}
\label{app:separability}

The main RDMs use condition means because the single-trial clusters overlap strongly. This conclusion follows from three measures: nearest-mean accuracy, tangent-space silhouette, and a geodesic scatter ratio. The silhouette remains near zero on every cohort, ranging from $-0.04$ in-house to 0.02 on BCIC, so individual trials do not form well-separated condition clusters.

Condition means nevertheless retain class information. Nearest-mean accuracy and scatter both rise from the in-house cohort to Weibo and BCIC: their values are 0.18 and 0.03 in-house, 0.39 and 0.16 on Weibo, and 0.55 and 0.19 on BCIC. Decoding therefore relies on distances to condition averages, which is precisely the structure captured by the condition RDM.

The SPDNet representation increases scatter by a factor of 1.4--1.6, but it reduces nearest-mean accuracy by 0.02--0.05 and leaves the silhouette near zero. Its dimension reduction removes within-class variation together with some class-relevant directions, consistent with the decoder pattern in \sectionref{sec:saturation}.

\newpage
\section{Measured agreement can exceed the true shared structure}
\label{app:ann}

The quantity $\rho$ estimates is the agreement between subjects' latent condition
configurations, and in EEG those configurations are never observed. An observed $\rho$
therefore cannot be compared against its target: when an alignment changes $\rho$, we cannot
tell whether it recovered structure the subjects share or imposed similarity that was not
there. This appendix constructs a synthetic setting in which the latent configurations are
generated explicitly, so the true agreement $\rho_{\mathrm{GT}}$ is known and the observed
$\rho$ can be checked against it. The setting is built to test how the measure behaves, not
to reproduce the EEG cohorts.

\paragraph{Setup.}
Twelve artificial domains stand in for subjects, each with six classes in
$\mathbb{R}^{16}$ playing the role of task conditions. Every domain gets its own version of
a shared class configuration, $\mathbf{M}_{s,c}=\mathbf{M}_c+\varepsilon_{s,c}$ with
$\varepsilon_{s,c}\sim\mathcal{N}(0,\sigma_{\mathrm{rel}}^2 \mathbf{I})$, plus a nuisance map
$x=\mathbf{A}_s z+b_s$ carrying domain identity. A small multilayer perceptron is trained on the
pooled data with class labels only. All measurements are taken on trials held-out from
training, using the activations of the layer before the classifier output. This layer is the
network's learned representation of a trial, so the RDM and the domain probe describe what
the network has learned rather than the synthetic features it was given.

Two regimes differ only in how much geometry the domains share. In the \emph{partly shared}
regime ($\sigma_{\mathrm{rel}}=2.5$) each domain arranges its classes somewhat differently,
and the true agreement is $\rho_{\mathrm{GT}}=0.30$. In the \emph{fully shared} regime
($\sigma_{\mathrm{rel}}=0$) all domains have the same arrangement, so
$\rho_{\mathrm{GT}}=1.00$. The fully shared regime shows the alignments can raise agreement
at all, without which a failure elsewhere would be uninterpretable.

RDMs use Euclidean distances between class means. Per-domain mean centering is an isometry
of that metric, the Euclidean counterpart of the AIRM congruence in
\theoremref{thm:invariance}. We compare it against two maps that are not isometries:
per-domain diagonal standardization (\emph{z-score}, low capacity) and per-domain
second-moment matching to a shared reference (\emph{coral}, high capacity). Estimators match
the main text, and the probe and RDM use disjoint test splits.

\begin{table}[h]
\centering
\caption{Artificial-network check, means over three seeds; probe chance $1/12=0.083$. A
positive $\rho{-}\mathrm{GT}$ means the measure reports more shared structure than exists.
$\max|\Delta \mathbf{D}|$ is the largest entrywise RDM change from the unaligned features.}
\label{tab:ann}
\begin{tabular}{llccccc}
\toprule
Regime & Variant & probe & $\rho$ & $\rho{-}\mathrm{GT}$ & $\kappa$ & $\max|\Delta \mathbf{D}|$ \\
\midrule
\multirow{4}{*}{Partly shared ($\rho_{\mathrm{GT}}{=}0.30$)}
 & none (unaligned)      & 0.921 & 0.461 & $+0.156$ & 0.626 & $0$ \\
 & per-domain center     & 0.194 & 0.461 & $+0.156$ & 0.626 & $3.6\times10^{-14}$ \\
 & z-score               & 0.221 & 0.462 & $+0.157$ & 0.579 & $47$ \\
 & coral                 & 0.038 & 0.837 & $\mathbf{+0.532}$ & 0.898 & $17$ \\
\midrule
\multirow{4}{*}{Fully shared ($\rho_{\mathrm{GT}}{=}1.00$)}
 & none (unaligned)      & 0.813 & 0.694 & $-0.306$ & 0.801 & $0$ \\
 & per-domain center     & 0.136 & 0.694 & $-0.306$ & 0.801 & $2.6\times10^{-14}$ \\
 & z-score               & 0.160 & 0.731 & $-0.269$ & 0.820 & $40$ \\
 & coral                 & 0.036 & 0.945 & $-0.055$ & 0.968 & $10$ \\
\bottomrule
\end{tabular}
\end{table}

\paragraph{An isometry removes identity but cannot move the measure.}
\tableref{tab:ann} reports the full set of runs. Mean centering drops the domain probe from 0.921 to 0.194 when geometry is partly shared,
and from 0.813 to 0.136 when it is fully shared. In both cases the RDM changes by at most
$3\times10^{-14}$, so $\rho$ and $\kappa$ do not move. This reproduces
\theoremref{thm:invariance} in an artificial network under a different metric, so the
dissociation seen in EEG is not specific to EEG, to AIRM, or to the in-house cohort. It
follows from how the alignment relates to the metric that evaluates it.

\paragraph{A high-capacity map pushes the measure past the truth.}
Coral removes identity in both regimes, but its effect on agreement depends on whether
shared structure exists. When geometry is fully shared, coral recovers it: $\rho$ rises from
0.694 to 0.945 against a truth of 1.00. When geometry is only partly shared, the same
operation raises $\rho$ from 0.461 to 0.837, overshooting the true 0.30 by 0.53. Coral did
not find correspondence that was there; it made the domains resemble each other by reshaping
them. Taken alone, an agreement of 0.837 would look like successful alignment, and it would
be wrong.
\paragraph{Why this matters.}
Alignment studies commonly read a change in $\rho$ as a change in shared structure. Both
results above break that reading, in opposite directions. Under an isometry, $\rho$ cannot
move regardless of how well the alignment works, so an unchanged $\rho$ is evidence about
the metric rather than about the alignment. Under a high-capacity map, $\rho$ can rise above
$\rho_{\mathrm{GT}}$, so an increased $\rho$ does not establish that the alignment recovered
shared structure. The noise ceiling does not separate these cases. Coral raises $\kappa$
from 0.626 to 0.898 at the same time as it raises $\rho$, so $\kappa-\rho$ falls from 0.165
to 0.061 while $\rho$ moves furthest above $\rho_{\mathrm{GT}}$. The smallest difference in
the table therefore belongs to the most inflated agreement estimate.

\sectionref{sec:training} shows the same failure of $\kappa-\rho$ from the opposite side.
There, increasing the dCKA penalty narrows the difference from 0.173 to 0.116, but $\rho$
falls from 0.153 to 0.135 while $\kappa$ falls from 0.326 to 0.252: the difference shrinks
because reliability collapses, not because agreement improves. Here it shrinks because
agreement climbs past a ceiling that rises more slowly. In both cases $\kappa-\rho$ moves in
the direction that would ordinarily be read as better alignment, while the underlying
quantity moves the wrong way or not at all. The EEG experiment can identify the confound but
cannot rule out the inflation case, because $\rho_{\mathrm{GT}}$ is unavailable there; this
setting separates the two. Together the two experiments show that neither $\rho$ nor $\kappa-\rho$ can be interpreted
without knowing which transformations the alignment can apply to the representation: an
alignment that cannot alter the RDM makes $\rho$ uninformative, and one flexible enough to
reshape it can drive $\rho$ away from $\rho_{\mathrm{GT}}$.

\end{document}